\documentclass[12pt,preprint]{aastex}

\usepackage{graphicx}
\usepackage{natbib}
\usepackage{amssymb}
\usepackage[figuresright]{rotating}

\newcommand{\gapprox}{$\stackrel {>}{_{\sim}}$}   
\newcommand{\lapprox}{$\stackrel {<}{_{\sim}}$}

 \slugcomment{To appear in Ap. J.}

\begin{document}

\title{Simultaneous monitoring of the photometric and polarimetric
activity of the young star PV Cep in the optical/near-infrared
bands
\thanks{Based on observations collected at AZT-24 telescope
(Campo Imperatore, Italy), AZT-8 (Crimea, Ukraine), and LX-200 (St.Petersburg, Russia)}}

\author{D.Lorenzetti\altaffilmark{1},
T.Giannini\altaffilmark{1},
V.M.Larionov\altaffilmark{2,3,4},
A.A.Arkharov\altaffilmark{3},
S.Antoniucci\altaffilmark{1},
A.Di Paola\altaffilmark{1},
T.S.Konstantinova\altaffilmark{2},
E.N.Kopatskaya\altaffilmark{2},
G.Li Causi\altaffilmark{1},
and
B.Nisini\altaffilmark{1}
}
\altaffiltext{1}{INAF - Osservatorio Astronomico di Roma, via
Frascati 33, 00040 Monte Porzio, Italy, lorenzetti, giannini,
antoniucci, dipaola, licausi, nisini@oa-roma.inaf.it}
\altaffiltext{2}{Astronomical Institute of St.Petersburg
University, Russia, vlar2@yandex.ru, azt8@mail.ru, enik1346@rambler.ru}
\altaffiltext{3}{Central Astronomical Observatory of Pulkovo,
Pulkovskoe shosse 65, 196140 St.Petersburg, Russia,
arkadi@arharov.ru}
\altaffiltext{4}{Isaac Newton Institute of Chile, St.Petersburg
branch}

\begin{abstract}
We present the results of a simultaneous monitoring, lasting more than 2 years,
of the optical and near-infrared photometric and polarimetric activity of the variable protostar PV Cep.
During the monitoring period, an outburst has occurred in all the photometric
bands, whose declining phase ($\Delta$J $\approx$ 3 mag) lasted
about 120 days. A time lag of $\sim$ 30 days between optical and infrared light curves has been measured and interpreted in the framework of an accretion event.
This latter is directly recognizable in the significant
variations of the near-infrared colors, that appear bluer in the outburst phase, when the star
dominates the emission, and redder in declining phase, when the disk emission prevails.\\
All the observational data have been combined to derive a coherent picture of the complex
morphology of the whole PV Cep system, that, in addition to the star and the accretion
disk, is composed also by a variable biconical nebula. In particular,
the mutual interaction between all these components is the cause of the high
value of the polarization ($\approx$ 20\%) and of its fluctuations. \\
The observational data concur to indicate that PV Cep is not a genuine EXor star, but rather a
more complex object; moreover the case of PV Cep leads to argue about the classification of other
recently discovered young sources in outburst, that have been considered, maybe over-simplifying,
as EXor.
\end{abstract}


\section{Introduction}

PV Cephei ($\alpha_{2000}$ = 20$^{h}$45$^{m}$53.96$^{s}$,
$\delta_{2000}$ +67$^{\circ}$57$^{\prime}$38.9$^{\prime \prime}$)
is a pre-main sequence star in the northeastern edge of the L1158
and L1155 group of dark clouds, at a distance of about 500 pc
(Cohen et al. 1981). It underwent significant short lived
outbursts detected in the optical and near IR bands, and,
therefore, it was classified as an EXor object (Herbig 1989). The
commonly accepted spectral type of PV Cep is A5 (see e.g.
\'{A}brah\'{a}m et al. 2000 and references therein), but, during
local minimum, it shows an absorption spectrum attributable to a
later spectral type G8-K0 (Magakian \& Movsesian 2001). Indeed,
its nature has been never certainly stated, although it has been
studied in several occasions and at different frequencies:
sometimes it has been classified as an Herbig Ae star (e.g. Fuente
et al. 2002), or as a FUor system. No doubt it is an embedded
young object whose phenomenology is partly attributable to
recurrent accretion from the circumstellar disk (Hartmann \&
Kenyon 1985). With this respect, the existence of an edge-on
circumstellar disk was indirectly suggested by the presence of ice
absorption in the 2-4 $\mu$m spectrum (Van Citters \& Smith 1989).
A more recent interferometric observation at 1.3 and 2.7 mm has
been able to resolve a relatively large (500 AU) and massive (0.8
M$_{\sun}$) disk (Hamidouche 2010).

According to a widely accepted picture, these systems accrete
material from their disks through rapid and intermittent events
that generate thermal instabilities in the disk itself and,
eventually, outbursts phenomena. Accreted matter migrates toward
the central star where it is channelled along the magnetic field
lines (e.g. Shu et al. 1994). The fall onto the stellar surface
produces a shock that cools by emitting a hot continuum; moreover,
as a consequence of the accretion event, strong winds (in some
cases also collimated jets) emerge from the rotating star/disk
system. Accretion disks survive few million years before
dissipating, thus it is plausible that accretion events go on for
a period shorter than that, with an outburst intensity that
decreases with time.

The interaction with a close binary companion is also invoked as
an alternative mechanism to produce accretion disk instabilities
and consequent outbursts (Clarke, Lin \& Pringle 1990; Bonnell \&
Bastien 1992), but recent IR surveys (even with adaptive optics)
for searching a binary close to PV Cep (Connelley, Reipurth \&
Tokunaga 2008, 2009) have given a negative result.

PV Cep is associated with GM-29 (RNO 125), an optically visible
(Cohen et al. 1977; Gyul'budagyan \& Magakyan 1977) and infrared
(Connelley, Reipurth \& Tokunaga 2007) variable nebula (with an
intermittent north-east streak) whose very rapid changes (both in brightness and in morphology)
are more likely due to a variable illumination from the star rather than to an intrinsic phenomenum.
Infrared analogs are the variable nebulae associated with both the
protostar L483 (Connelly, Hodapp, \& Fuller 2009) and the eruptive
object V1647 Ori (Brice\~{n}o et al. 2004; Fedele et al. 2007).

PV Cep drives the giant HH 315 flow (G\'{o}mez,
Kenyon \& Whitney 1997; Reipurth, Bally \& Devine 1977) and a
molecular outflow (Levreault 1984, 1988), both aligned with the
symmetry axis of the nebula at a position angle of 330$^{\circ}$. The
blue-shifted $^{13}$CO integrated emission shows a V-shaped morphology
coincident with the reflection nebula, while the blue-shifted $^{12}$CO
traces fan-like morphology that fills the cavity delineated by the
$^{13}$CO structure (Arce \& Goodman 2002). These latter authors
suggest that the $^{13}$CO traces the limb-brightened walls of a
wide-angle wind blown cavity.

Large values of intrinsic polarization ($\approx$ 10-15 \%) and
evidence of polarimetric variability were detected by Bastien \&
M\'{e}nard (1988) and M\'{e}nard \& Bastien (1992). In the period
1981-1989, polarization maps were obtained by Gledhill et al.
(1987) and Scarrott et al. (1991), who proposed a model for the PV
Cep/GM29 system according which: {\it (i)} the main characteristic
of the system is its rapid brightness and structural variability,
sometimes on the time-scale of months; {\it (ii)} the intermittent
north-east streak is a reflection effect caused by sporadic
illumination of the outer parts of the cavity wall by radiation
from the central star; {\it (iii)} the star is seen through at
least two sets of polarizing grains with different alignments and
the relative amount of extinction by each set has changed with
time.

We have started our monitoring of PV Cep since some years, and
part of the data have been already presented in Larionov et al.
(2007) and  Arkharov et al. (2008). The present paper deals with a
more than 2 years, multi-wavelength monitoring (visual, near-IR
photometry and I-band polarimetry), during which PV Cep underwent
an outburst followed by a very rapid and significant declining;
our aim is put the complex phenomenology
of PV Cep into a coherent scheme. \\
Quite recently, the results of an optical (V,R,I) monitoring of PV
Cep have been presented (Kun et al. 2011) complemented with some
{\it Spitzer} mid-IR photometric data. Their optical light-curves
present an overlap with ours although they cover a longer period:
within the overlapping region their data are fully in agreement
with those presented here; near-IR data are substantially our data
retrieved from literature. Kun et al. (2011) suggest that the
photometric decline they have sampled resulted from an interplay
between variable accretion and circumstellar extinction. A
comparative analysis between their and our conclusions will be
done in the following
(Sects. 3.1.2, 3,3, 3.4 and 4). \\
The paper is structured as follows: in Sect.2 the observations are
presented, while in Sect.3 the results are provided in the context
of the previous knowledge of this source. Finally, our conclusions
are summarized in Sect.4.

\section{Observations}

All the observations (see also Table~\ref{summary:tab}
in Appendix) were obtained during the period April 2007 - Nov.
2009. Noticeably, the intra-day or day time-scale variations are
not relevant for the discussion, so that observations in different bands
have been associated to the same date (within a maximum distance of 1 day)
even if not strictly simultaneous.

\subsection{Optical photometry and polarimetry}

Optical photometric and polarimetric observations were obtained
with two nearly-identical photometers-polarimeters of the
Astronomical Institute of St.Petersburg University, mounted on the
70cm AZT-8 telescope of the Crimean Observatory (Ukraine) and 40cm
LX-200 telescope of St.Petersburg University, respectively. These
photometers are based on ST-7XME SBIG CCDs and equipped with UBV
(Johnson) and RI (Cousins) filters. In polarimetric mode two
Savart plates are used as analyzers, giving $q$ and $u$ Stokes
parameters. AZT-8 observations were obtained in R and I bands,
while LX-200 ones in unfiltered mode, whose effective wavelength
roughly corresponds to the R band. The pixel scale is 0.64
arcsec/pxl which corresponds to a 8.1$\times$5.4 arcmin$^2$ field
of view.

The optical light curves are given in the three upper panels of
Figure~\ref{IRmag:fig} for the V, R, and I band, respectively, while the
polarimetric results are listed in Table~\ref{summary:tab} of
the Appendix.

\subsection{Near-IR photometry}

Near-IR data were obtained at the 1.1m AZT-24 telescope located at
Campo Imperatore (L'Aquila - Italy) equipped with the
imager/spectrometer SWIRCAM (D'Alessio et al. 2000), which is
based on a 256$\times$256 HgCdTe PICNIC array. Photometry is
performed with broad band filters J (1.25 $\mu$m), H (1.65
$\mu$m), and K (2.20 $\mu$m). The total field of view is
4.4$\times$4.4 arcmin$^2$, which corresponds to a plate scale of
1.04 arcsec/pixel. All the observations were obtained by dithering
the telescope around the pointed position. The raw imaging data
were reduced by using standard procedures for bad pixel removal,
flat fielding, and sky subtraction. The derived light curve is
depicted in the three lower panels of Figure~\ref{IRmag:fig} for
the J, H, and K band, respectively.

\section{Results and discussion}

\subsection{Amplitude and colors variations}

\subsubsection{Near-infrared colors}

PV Cep light-curves presented in Figure~\ref{IRmag:fig} show how
the source has undergone a significant variability during the
monitoring period of more than 2 yr. In particular a rapid ($\sim$
120 days) and significant ($\Delta$J $\ga$ 3 mag) declining is
evident between MJD 54558 and 54673 (even if with a short rising
lasting about 10 days).\

\begin{figure}
\includegraphics[angle=0,width=15cm] {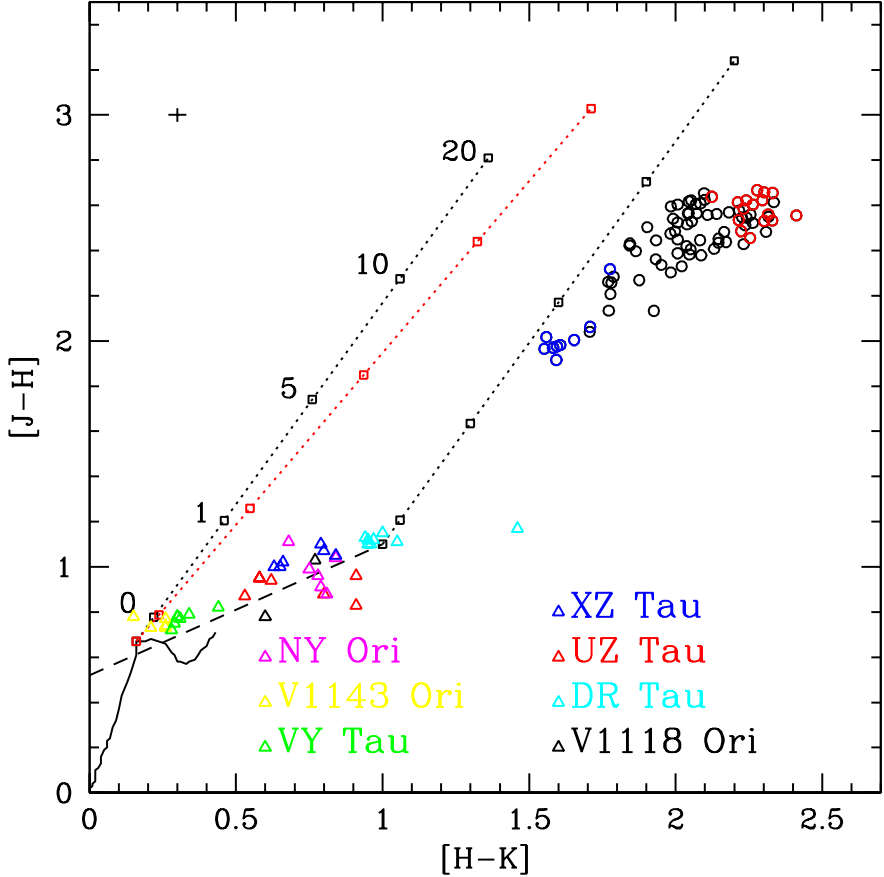}
   \caption{Near-IR two colors diagram of PV Cep in different epochs.
   The solid line marks the unreddened main sequence, whereas the dashed
   one is the locus of the T Tauri stars (Meyer et al. 1997). Black (red)
   dotted lines represent the reddening law by Rieke \& Lebofsky (1985)
   and Cardelli et al. (1989), respectively, where different values of A$_V$ are indicated by open
   squares. Blue (red) dots indicate those cases in which the source was found in its highest
   (lowest) state, arbitrarily identified with a J-band magnitude lesser (greater) than 11.0
   (12.5). The near-IR colors of EXor stars (Lorenzetti et al. 2009),
   that are targets of our on-going monitoring program, are also
   depicted with triangles and evidently overlap with the locus typical of T Tauri stars.
   A cross in the upper left corner indicates the typical error.
   \label{IRcolcol:fig}}
\end{figure}

Aiming at investigating the possible origin of the photometric
variations, we plot in Figure~\ref{IRcolcol:fig} the near-IR [J-H]
vs [H-K] color diagram of PV Cep. All the data cluster in the red
portion of the plot, but important differences exist. To indicate
the star bright and faint states we have marked with blue (red)
dots the points with J-magnitudes lesser (greater) than 11.0
(12.5), while black dots indicate J-magnitudes between these two.
From the diagram it is quite evident the PV Cep becomes bluer
(redder) while brightening (fading). This variability cannot be
explained just with an extinction variation: in the same Figure we
have depicted two different extinction laws (Rieke \& Lebfosky
1989 and Cardelli et al. 1989): the points are not aligned with
any extinction curve, but exhibit a clear shift to the right, that
increases with decreasing the PV Cep brightness.

We have searched  also for different mechanisms as a possible origin
of the near-IR colors deviation. The scattering contribution has
been evaluated following the relationships (2) and (3) given in
Massi et al. (1999), valid for optically thin ($\tau_0 <$ 1) dust
producing an isotropic Rayleigh scattering. The scattering
contribution for values of $\tau_0$ between 0.1 and 1 produces
color changes of just one tenth of magnitude. Finally, we have
considered the effect of a hot spot on the stellar surface. As
example we have assumed a hot spot temperature of T=12000K and a
surface equal to 40\% of the stellar one, finding just negligible
changes of near-IR colors (few hundredths of magnitudes).

Summarizing, from this part of the analysis we can deduce that:
{\it i)} the color variations do not follow the extinction vector
and are thus not due varying extinction, while the colors are
reddened by a roughly constant extinction A$_V$ of about 5-7 mag;
{\it ii)} the range of the color variations is similar in both
colors ($\sim$ 0.8 mag); {\it iii)} the observed variations
reflect an intrinsic phenomenum, likely due to the circumstance
that, when outbursting, the central star becomes more visible,
while when declining, the disk dominates; {\it iv)} the observed
variations cannot be accounted for just by extinction
fluctuations; other phenomena, such as scattering variations
and/or hot spots on the stellar surface give negligible
contributions.

\subsubsection{Optical colors}

\begin{figure*}
\includegraphics[angle=0,width=15cm] {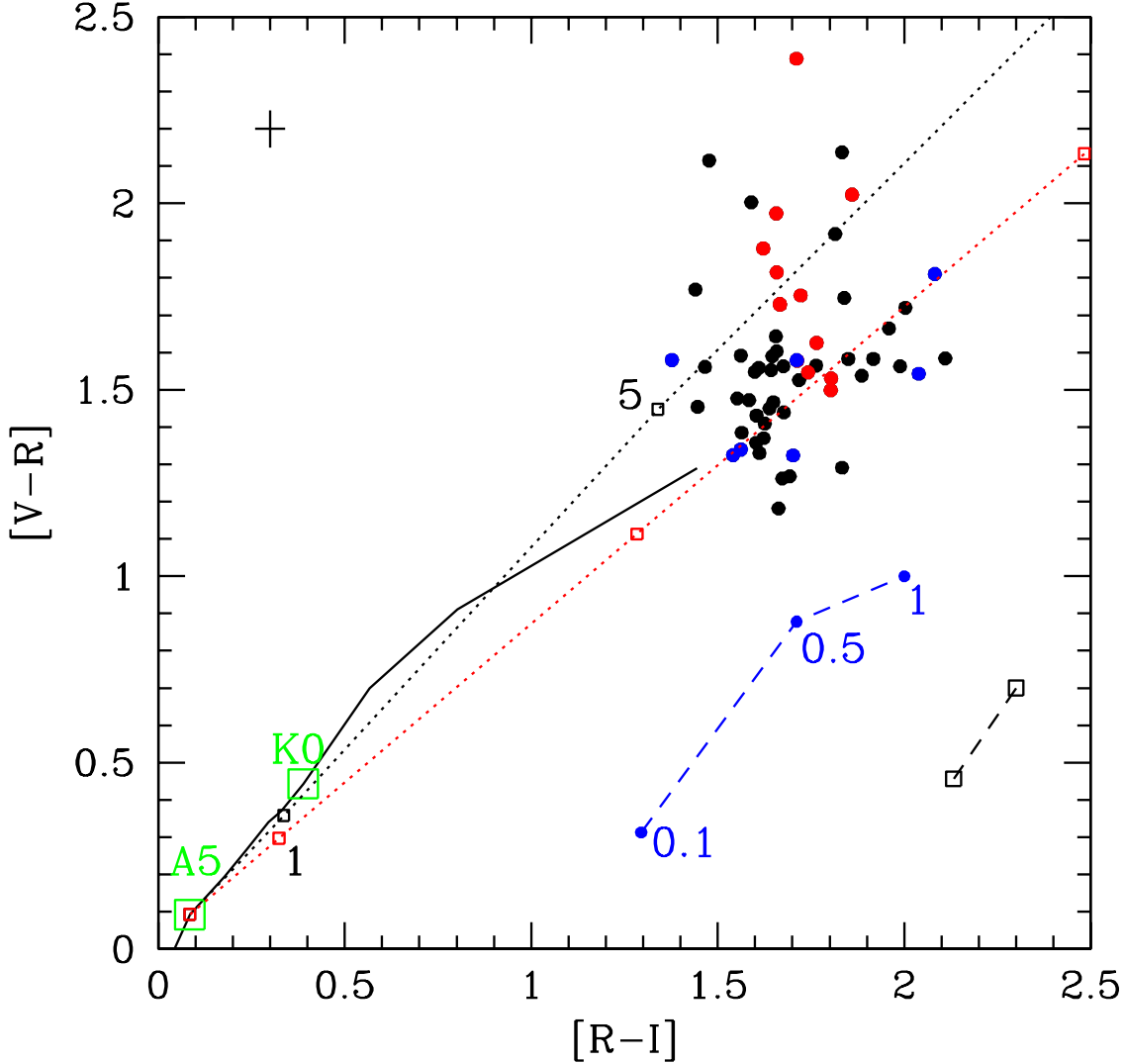}
   \caption{Visual two colors diagram of PV Cep in different epochs.
   The solid line marks the unreddened main sequence star of a given
   spectral type as defined by Kenyon \& Hartmann (1995) and corrected from Johnson
   to Cousins system (Fernie 1983); the relevant (see text)
   spectral types A5 and K0 are indicated as green open squares.
   The black (red) dotted lines represent the reddening laws by
   Rieke \& Lebofsky (1985) and Cardelli et al. (1989), respectively,  where different
   values of A$_V$ are indicated by open squares (0, 1, 5 mag).
   The blue curve arbitrarily applied on the bottom right represents the color variations shown
   by a given star for different values of the optical depth ($\tau_0$) of the scattering
   nebula (from 1 to 0.1). By decreasing $\tau_0$ (from 1 to 0.1) the scattering contribution
   increases and the star appears bluer. Blue (red) dots indicate those cases in which
   the source was found in its highest (lowest) state, arbitrarily
   identified with a V-band magnitude lesser (greater) than 17.7
   (18.7). In the bottom right corner the blueing effect of
   adding a hot spot (having T = 12000 K and a surface equal to 40\% of the
   stellar one) to an arbitrary point, is quantitatively indicated with squares.
   Finally, a cross in the upper left corner indicates the typical
   error.
   \label{VRIcol:fig}}
\end{figure*}

It is worthwhile to repeat the above analysis by using the optical
colors V, R, and I, where extinction and scattering are both
expected to play a major role (Figure~\ref{VRIcol:fig}). In
analogy with the near-infrared diagram, we indicate with blue
(red) dots the points corresponding to high (low) brightness,
identified with a V-band magnitude lesser (greater) than 17.7
(18.7). All data points roughly cluster in a locus corresponding
to a main sequence A5 (K0) star extincted by 6-7 (5-6) mag: this
is what suggested also by near-IR colors and exactly what is
expected for PV Cep (see Sect.1).

The well defined color segregation shown by PV Cep in the near-IR,
appears here barely recognizable. We ascribe this difference to a
combination of two facts: ({\it i}) the reduced effect (at visual
wavelengths) of the intrinsic IR excess;  ({\it ii}) the presence
of additional contributions that, while negligible in IR, become
important in the optical. For example (Figure~\ref{VRIcol:fig},
bottom right corner), the effect of a variable scattering (with
$\tau_0$ values from 0.1 to 1), gives color fluctuations
compatible with a variable scattering nebula. Similarly, the
presence of a hot spot on the star surface, with the same physical
characteristics as those considered in the near-IR, gives
substantial variations of about 0.5 mag in both colors.
Remarkably, the optical colors are completely inconsistent with
A$_V$ variations in the range 7-15 mag, thus confirming what
suggested by the near-IR plot. This comparison rules out the
occurrence of large A$_V$ changes and further demonstrates how the
near-IR color variations cannot be due to extinction variations,
but are originated by an intrinsic excess fluctuations.

Summarizing, from the VRI two colors plot we can deduce that: {\it i)}
irrespectively of the adopted extinction law, A$_V$ could be
allowed to vary only in the range 6-8 mag; {\it ii)} other phenomena (than
extinction) may have a significant role; {\it iii)} the range of the [V-R]
color variations ($>$ 1 mag) is significantly larger than that of
[R-I] variations ($\sim$ 0.5 mag).

A different conclusion, in favour of an A$_V$ component of about 3
mag, has been reached by Kun et al. (2011) through the analysis of
their light-curves. This discrepancy may be due to the lack of
simultaneity between their optical and near-IR photometry. Indeed,
it is just the comparative analysis between optical and near-IR
color plot that allowed us to rule out any A$_V$ variation during
our monitoring period. In our opinion, this circumstance
represents a clear example of how a simultaneous and
multi-frequency monitoring of several brightening and fading
events will lead to a more comprehensive conclusion, ie. not too
related to the nature of single events.




\subsection{Time delay in the light curve}

\begin{figure}
\includegraphics[angle=0,width=15cm] {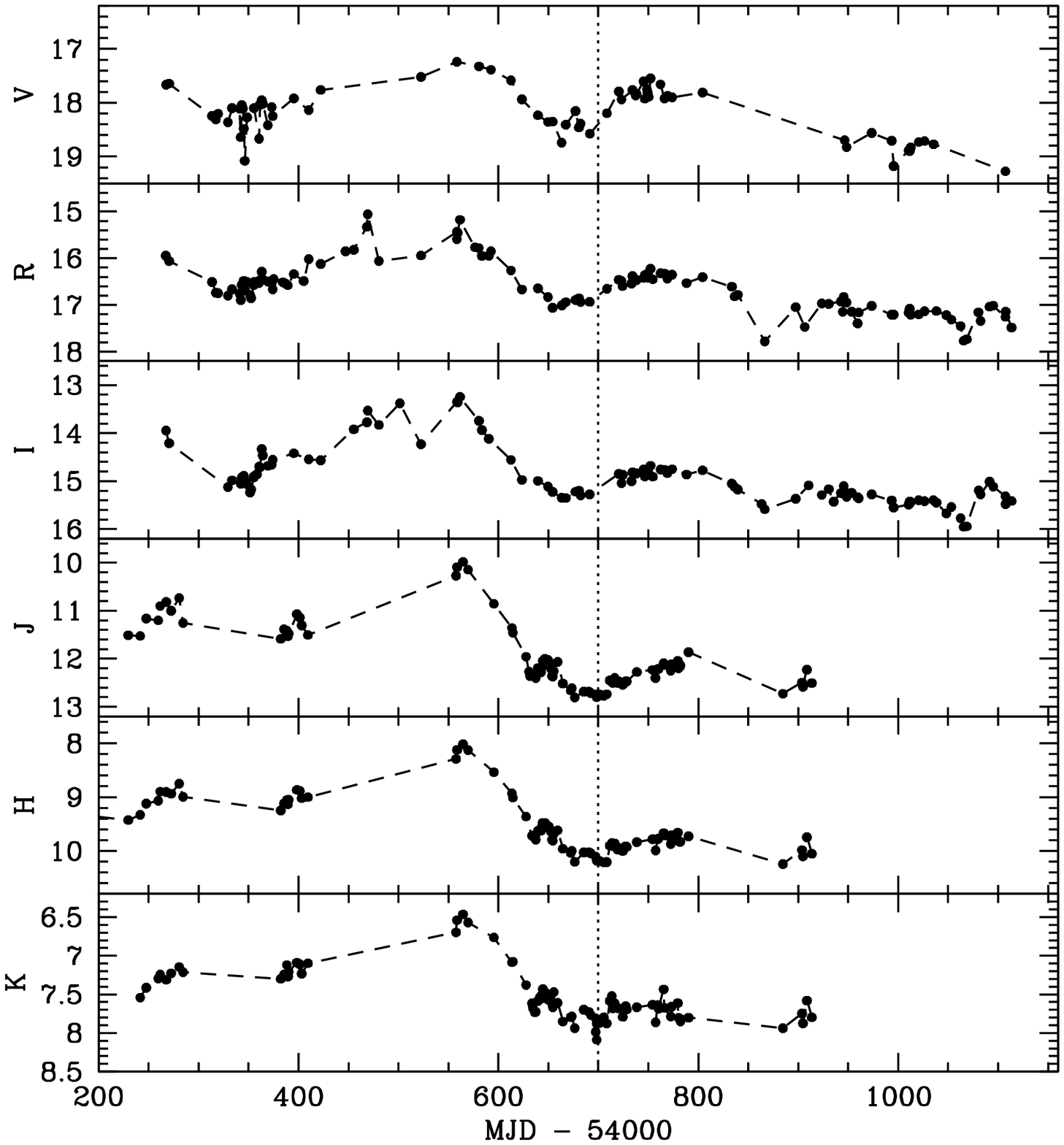}
   \caption{PV Cep optical and near-IR light curves vs. MJD (Modified Julian Date).
   The errors of data points are comparable to or lesser than 0.02 mag.
   The black vertical dotted line is plotted to have a temporal reference for a by-eye estimate
   of possible lags at different wavelengths.
   \label{IRmag:fig}}
\end{figure}

\begin{figure*}
\includegraphics[angle=0,width=15cm] {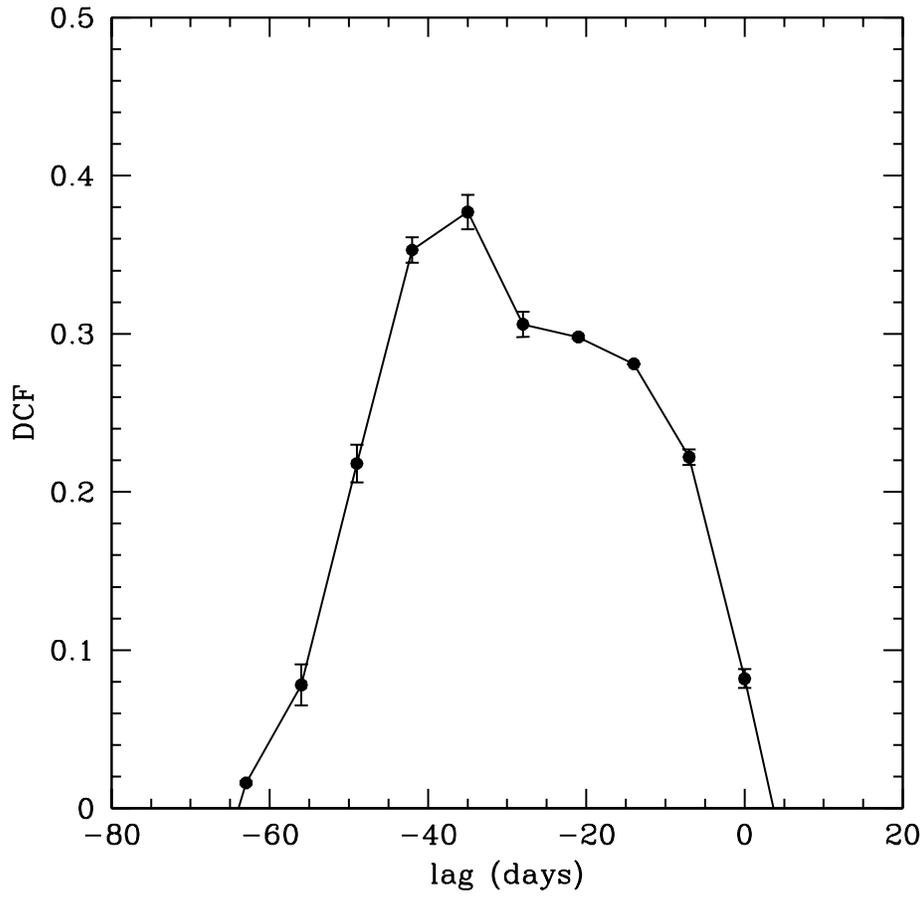}
   \caption{Discrete Correlation Function (DCF) for PV Cep optical
   ($R$ band) and near-IR ($J$ band).
   \label{dcfCep:fig}}
\end{figure*}

By examining the light curves in different bands
(Figure~\ref{IRmag:fig}) some delay seems to exist between optical
and near-IR observations, namely any event (characterized by an increasing or
decreasing flux) appears to take place before at shorter
wavelengths (see the black vertical dotted line in
Figure~\ref{IRmag:fig}). However, the possible existence of a temporal lag
has to be supported in a more quantitative way. The common way to
estimate delays between two light curves at different wavelengths
is to use the Discrete Correlation Function (DCF - Edelson \&
Krolik, 1988). This method is suitable for measuring
correlation functions when the sampling time is difficult to
control (e.g. objects observed in different bands at different phases).
The robustness of the DCF has been demonstrated by Hufnagel \& Bregman (1992), who pointed out
how the shape of the DCF reflects the underlying nature of the process to which the data refer.
The parameters of the DCF are the width, that indicates the duration of the correlated phenomena
between the two data sets, and the peak height, that is a measure of the correlation strength.
To estimate delays between optical ($R$ band) and near-IR
($J$ band) we used the data in the time interval from MJD~545500 to
MJD~54750, most densely covered in both bands. Our computation indicates
(at a DCF level of about 40\%, see Figure~\ref{dcfCep:fig}) that variations in the $J$
band occur around 30 days after those in the $R$ band.

The existence of a lag between optical and near-IR light-curves
can be proved also in other YSO's. For example, Audard et al.
(2010) have qualitatively recognized some lag between the
near-infrared and optical light curves of the EXor variable V1118
Ori: we have applied the DCF method to their data (their Table 2)
obtaining the same lag of about 30 days, with a larger DCF peak
\lapprox 0.7.
The similarity of the lag duration speaks in favor of a common mechanism responsible
for these phenomena. Likely, we are simply looking at the response
of the emitting matter at each frequency (regulated by its thermal
capacitance) when it is undergoing an heating by outburst or a cooling.
If this is the case, the practical constance
of this lag should be a characteristics typical of any outbursting
or declining event of any disk-accreting protostar: to this end, a
careful investigation of the available optical/near-IR light
curves should be desirable. We remark that the lag occurrence is not in agreement
with an extrinsic variation eg. due to a circumstellar or
interstellar dust screen (such as proposed by Kun et al. 2011), which would generate strictly
simultaneous fluctuations at different frequencies. Conversely, the lag supports the
occurrence of an intrinsic phenomenum on the stellar surface (such
as an accretion event or an irregular hot spot) that requires,
because of its nature, a longer time to reach the maximum emission
at longer wavelengths.

\subsection{Polarization}


The PV Cep polarization (P) data collected  before our monitoring
are listed in Table~\ref{polar:tab}. Although these data stem from
a random sampling obtained at different epochs (from 1980 to 1989)
and in different bands, they indicate the large spread of observed
values on short time-scales, and a long term increasing (on years
time-scale) of the P maximum value. These qualitative indications
have motivated us to systematically monitoring the polarization
changes in PV Cep. To these new data (Table~\ref{summary:tab})
refer the plots of Figure~\ref{polar:fig}. Here the observed
values of the polarization, its position angle (PA) along with the
position angle of the main axis of the nebula (see below) are
given as a function of time. The R (I) band polarization and PA
values are marked as red (black) dots, while green dots represent
polarization and PA values obtained in unfiltered mode. In the
same plot the R-band light curve is also reproduced to make easy
the comparison with the PV Cep brightness state.

We first note that the mean value of the polarization
corresponding to our monitoring program is 19.5 \%. This high P
value is typical of a system viewed edge-on, like PV Cep: most of
the light from the star is attenuated and the fraction of the
scattered, polarized light, is high. Single or multiple scattering
in optically equatorial thick disks and/or in the lobes of the
associated bipolar nebulae are the widely accepted models to
produce patterns of aligned polarization vector in YSO's (Bastien
\& Menard 1988, 1990). The polarization maps of PV Cep (Gledhill
et al. 1987; Scarrott et al. 1991) also agree with the proposed
interpretation and their conclusions have been anticipated in
Sect.1. However, whichever kind of scattering in disks cannot produce
polarization values much larger than 12\%, therefore single
scattering in inhomogeneous dust distributions (such as bipolar
nebulae) is additionally required (Daniel 1982).

Now we aim at understanding whether or not the observed
polarization variations are
statistically significant and, eventually, at
inferring on the plausible causes of such variability.

To evaluate if polarization can be declared variable (within the
errors), we adopt the prescription given by M\'{e}nard \& Bastien
(1992), namely a 95\% confidence level established with a $\chi^2$
test of the Stokes parameters within a given pass-band.
The resulting P($\chi^2$) in our case is $>$ 99.9 \%, hence PV Cep
can be considered a polarimetric variable. This result is well in
agreement with the conclusion of the previous section 3.1 in favor
of a varying scattering (i.e. polarizing) nebula.

No clear correlation can be found between the polarization and
either the brightness in all the observed photometric bands or the
position angle of the polarization itself (PA, see
Figure~\ref{polar:fig}). These circumstances allow us to rule out
some models that predict the origin of the polarimetric variations
as uniquely due to accretion processes. Wood et al. (1996)
investigated the photo-polarimetric variability of a magnetic
accretion disk model for pre-main-sequence T Tauri stars. They find that
matter from the disk accretes along the magnetic field
lines onto the stellar surface producing hot and polarized spots;
stellar rotation causes the photometric and polarimetric
variations. However, at variance with our observational data, such model
predicts a correlation between brightness and polarization
variations, these latter occurring on time-scales comparable
to the stellar rotation period.

\begin{figure*}
\includegraphics[angle=0,width=15cm] {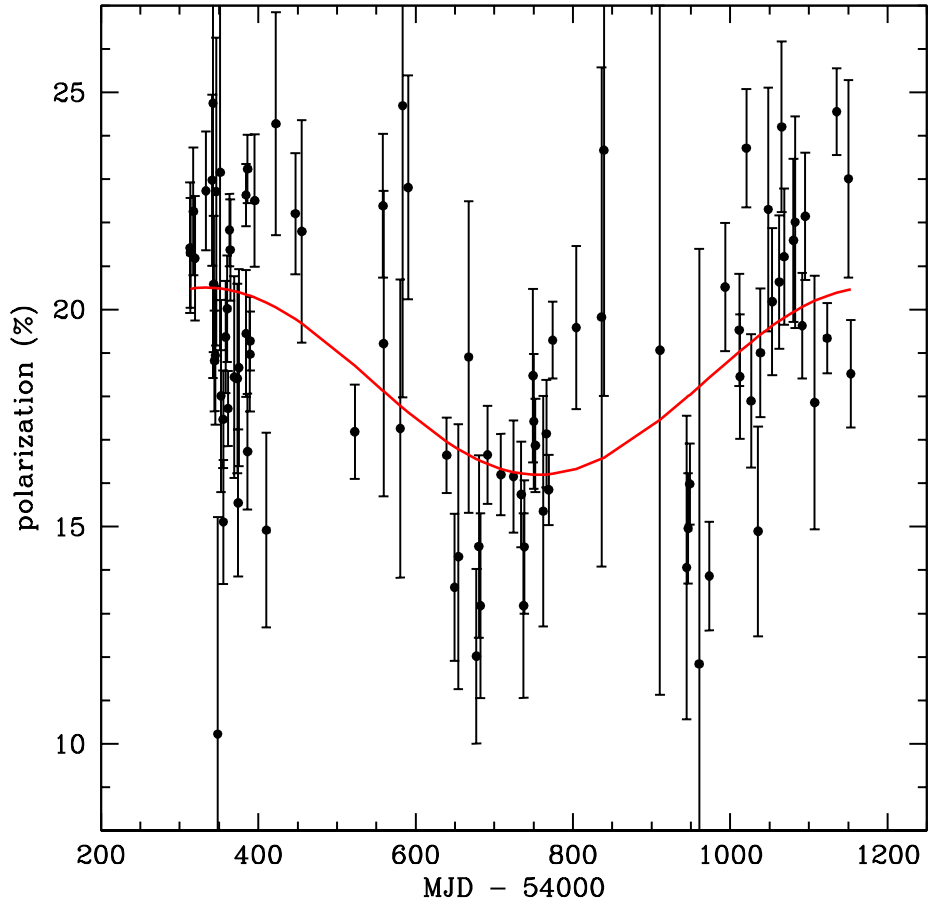}
   \caption{Temporal distribution of the PV Cep polarization intensity
   A sinusoidal fit to the data with period of 847 days and peak amplitude
   2.1\% is plotted, as well (red curve).
   \label{sinus:fig}}
\end{figure*}

We also searched for some periodicity in the polarization
variability, with indications of a positive result. To this aim we
used the complete set of polarization observations, that spans for
more than 2 years (Figure~\ref{sinus:fig}). The fit to the data
points with a sinusoidal behaviour gives formally  a period of 847
days and a peak amplitude of 2.1\% with a reduced $\chi^2$ = 2.7.
Hence, just one period of a pure sinusoid marginally (within
3$\sigma$) accounts for the polarization variability, whose
periodicity is somehow presumable. A significant increase of the
monitoring time is needed to confirm such preliminary indication,
also considering that non-periodic variations of different origin
are expected to occur. If this tentative interpretation were
confirmed, it would suggest a period typical of a disk orbital
motion or of a circumstellar envelope externally illuminated and
arranged in an asymmetric geometry. Also a circumbinary envelope
should be able to produce periodic polarization variations, but at
a very reduced amplitude, less than 1\% (Manset \& Bastien 2001).

The variations of PA, although not correlated with those of the
polarization amplitude, do not seem negligible, and show some
regularity. We can speculate, they also could be related to the
changes in geometry of the circumstellar material more than to
density inhomogeneities or clumps in the thick disk. This latter
could indeed produce apparently random variations of both the
polarization and its position angle (M\'{e}nard \& Bastien 1992).

By following Scarrott et al. (1991), we conclude that the strong
values of both the mean value of the
polarization and its variations could be due to the presence of at
least two populations of polarizing grains, having different
alignments and giving relatively different contributions. A fixed
and relevant amount of polarization could be due to the thickness of the
disk, and a variable (perhaps periodic) component could sit on top
of it, generated by an asymmetrically organized circumstellar
material.

\subsection{The nebula}

\begin{figure*}
\includegraphics[angle=0,width=15cm] {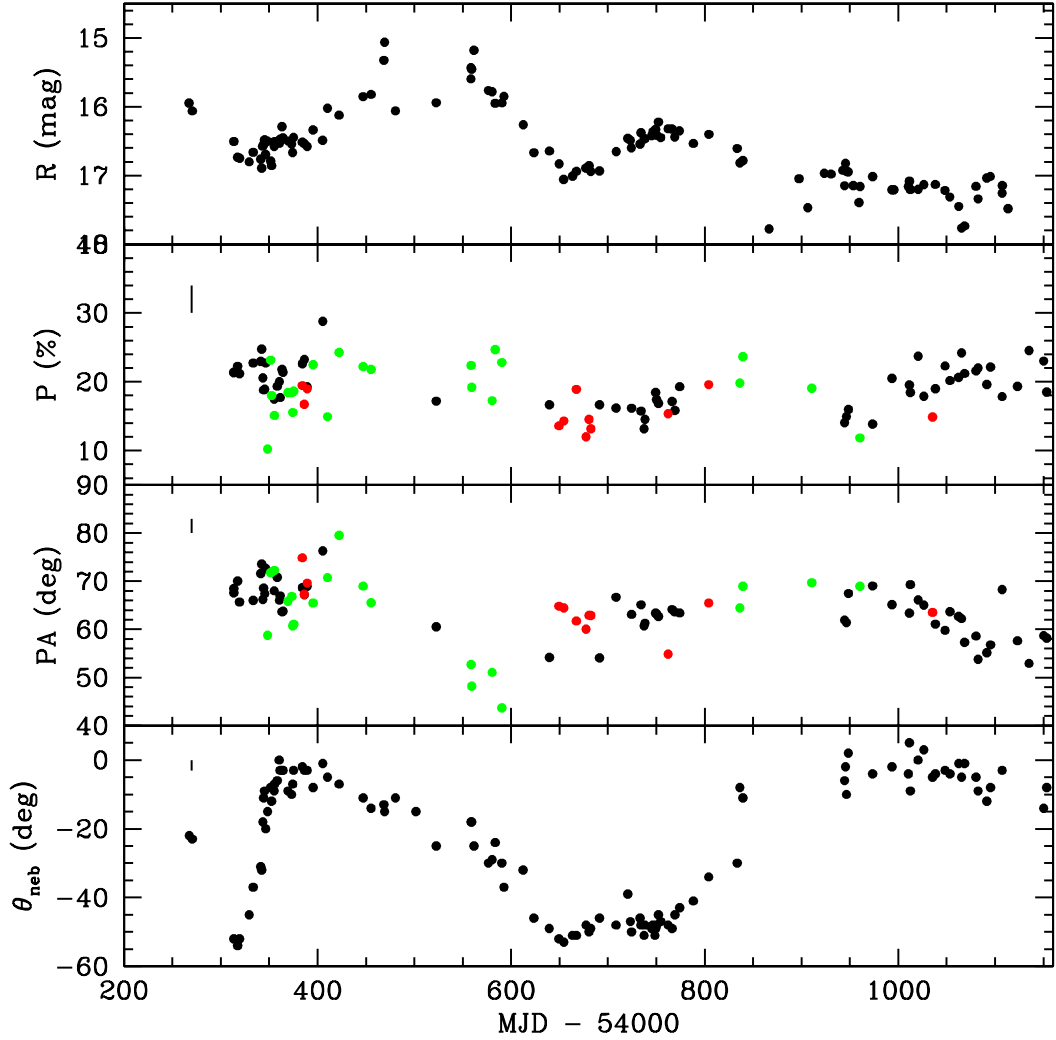}
   \caption{Temporal distribution of the PV Cep polarization intensity (in \%, second panel)
   and position angle (PA in deg, third panel), along with the direction of the cometary nebula
   main axis ($\theta_{neb}$, fourth panel). $R$ ($I$) band polarization and PA values are
   depicted as red (black) dots, while green dots represent polarization and PA values obtained
   in unfiltered mode. The light-curve in the $R$ band is plotted again (first panel),
   just for convenience. Typical errors are indicated by vertical
   bars in the upper left corner of each panel.
   \label{polar:fig}}
\end{figure*}

A photometric analysis of the (cometary) nebula associated to PV
Cep is not a task afforded in the present paper. Here the aim is
to put into light some phenomenology associated to the nebula that
could contribute to form a more coherent picture of the overall PV
Cep system. Our optical observations confirm that the northern
part of the nebula (ie. the one associated to the blue CO lobe) is by
far the brightest, while the southern one remains visible only in
few periods. Moreover, the nebula brightness tends to rather
closely correlate with the brightness of the object itself,
confirming its illuminated nature. The time fluctuations in
position of the nebula's main axis $\theta_{neb}$ (Figure~\ref{polar:fig},
bottom panel) are noticeable in the
present context. This angle is computed (from N to E) as the position angle of the
brightest portion of the fan, with a typical error of about 1-2$^{\circ}$.
In most cases, the brightest portion of the fan
coincides with the nebula axis, although sometimes it is difficult
to distinguish between the axis and the brightest blob. The
computed values of $\theta_{neb}$ are given in the last column of
Table~\ref{summary:tab}. For a typical system composed by a
bi-conical nebula emerging from an edge-on disk, the nebula axis
is expected to be orthogonal to the disk itself; indeed, being the
PV Cep disk on average oriented at 60$^{\circ}$ (see PA in
Figure~\ref{polar:fig}), the nebula's axis presents a
perpendicular average value of -30$^{\circ}$, but showing huge and
quite regular fluctuations of $\pm$30$^{\circ}$.

Firstly, we can exclude that this sort of nebula's axis precession
(or brightest blob movement) is due to a changing illumination
directly from the star (or from the disk), otherwise both would
present similar fluctuations with a fixed delay determined by the
light travelling time from the star to the blob ($\approx$ 30
days). Suitable tests (DCF analysis) done to search for possible
correlations provided negative results.

Instead, the nebula axis movement speaks in favor of some motion
of crucial parts of the overall structure. We can hardly believe
that the nebula or the disk are physically precessing in a so
short time-scales, rather we are seeing a permanent movement of
the hole (namely the innermost region of the disk itself) in the
CS disk through which the light penetrates to illuminate the fan.
Different portions of the walls delimiting the hole can plausibly
reflect not uniformly the starlight, then originating systematic
differences in the nebula illumination (see also Cohen et al.
1981). Such a mechanism is compatible with the time-scale of the
observed $\theta_{neb}$ fluctuation that are of the order of 1-2
yrs; however our data do not cover a temporal range long enough to
infer on a possible periodicity. Conversely, if the central star
were the direct illuminator, the time-scale of $\theta_{neb}$
fluctuation should be reconcilable with much shorter time-scales
(days) typical of the stellar dynamics (eg. rotation). According
to the proposed scenario, $\theta_{neb}$ (ie. the orientation of
the brightest part of the nebula) and PA (related to the grain
orientation) are expected to be rather uncorrelated (as confirmed
by Figure~\ref{polar:fig}. The proposed view agrees with
the interpretation given by Kun et al. (2011), according to which
the variable shape of the nebula is not only due to the fading of
the star, but also due to the changing geometry of the dust distribution
close to the star.

\subsection{Comparison with the EXor objects}

At this stage, after having derived some properties by our optical
and near-IR observations, we can come back to the issue of the PV
Cep classification (as Herbig Ae star or EXor system). It presents
many differences with respect to the classical EXor like XZ Tau,
UZ Tau, VY Tau, DR Tau, V1118 Ori, NY Ori, V1143 Ori, EX Lup
(Herbig 1989, Lorenzetti et al. 2009). In the light of the
presented data, the most relevant are: {\it i)} the bursts
modality evidenced in the light curve, that does not present a
rapid (weeks) increasing and slower (months /years) declining;
{\it ii)} the existence of a circumstellar nebulosity (with a
significant variability both in morphology and brightness), which
is not a so common feature for EXor's; {\it iii)} the high value
of the foreground extinction; {\it iv)} the near-IR colors (see
Figure~\ref{IRcolcol:fig}) that show larger fluctuations than
those exhibited by other Exor's. Other differences stem from
considerations presented in the literature: {\it i)} the
relatively early spectral type usually attributed to PV Cep and
its bolometric luminosity (about 100 L$_{\odot}$ instead of a few
or fractional solar luminosity; {\it ii)} the association with
mass loss manifestation (in terms of CO outflow and HH jets)
usually not detected in classical EXor's (Lorenzetti et al. 2006,
2007); {\it iii)} the presence of a rather massive circumstellar
material of $\approx$ 1 M$_{\odot}$ (Osterloh \& Beckwith 1995;
Fuente et al. 1998); {\it iv)} the presence of a significant
far-IR counterpart that dominates its luminosity (\'{A}brah\'{a}m
et al. 2000); {\it v)} the association with a radio continuum
likely arising in the shocked-ionized gas (Anglada et al. 1992);
{\it vi)} the presence of an H$_2$O maser source, although at a
relatively large distance ($\approx$ 3 10$^{16}$ cm) from the
visible star (Rodr\'{i}guez et al. 1987; Sunada et al. 2007).

Conversely, the similarity with other EXor stems from {\it i)}
being an emission line object; {\it ii)} interpreting the recurrent
bursts as mainly due to disk accretion events. \\

Considering all the presented evidences it seems that PV Cep is indeed an
object more massive and more complex than other EXor. It represents an example
suggesting that the EXor classification is not longer
adequately accounted for, especially in the context of the recent
detections of outbursting embedded objects, that have been often referred, maybe
over-simplifying, as EXor.

Most of these objects (such as V1647 Ori, OO Ser, EC 53, ISO ChaI
92, GM Cep) are indeed more embedded (i.e. younger) than classical
EXor. In principle, there is no conceptual reason which prevents
the EXor phenomena to occur in a more embedded phase. This implies,
however, that the well defined outburst modalities typical of visible, quite
isolated, low mass objects, become unavoidably less recognizable.
Consequently, the EXor class could include objects of different
nature. If EXor were uniquely recognized to be YSO's undergoing
random accretion phenomena (of some relevance in brightness variations),
then almost all the YSO's (Class I objects, Herbig Ae/Be stars, active T Tauri stars)
should be incorporated in that class, since they all present a
random and significant variability.

\section{Concluding remarks: a consistent picture}

We have presented the results of a simultaneous monitoring,
lasting more than 2 years, of the photometric and polarimetric activity of the
young stellar object PV Cep, in the optical/near-IR bands. This work aimed at
investigating its physical properties, the causes of its variability and the
nature of this object. By analyzing the results, here we try to delineate
a consistent picture of the PV Cep system.

PV Cep is a young Ae star with a quite massive circumstellar thick disk
associated to a bi-conical and variable (both in brightness
and in main axis direction) nebula.
While the disk rotates (in a typical time-scale of 1-2
years) it could present a not homogeneous internal edge to the
star light which, consequently, illuminates with a different
intensity different portions of the nebula. Reasonable
consequences of the mechanism described above are the recurrent
brightness variations of the nebula along with those of its main
axis, whose differences are also related to the disk rotation. The
strong mean value on the polarization is due to the thick disk,
while the variable (maybe periodic) component is related to the
different illumination of the nebula itself.

In addition to these phenomena, recurrent disk accretion events
have a role in producing intermittent outbursts in the light
curve. By examining the light-curves given in Figure~\ref{IRmag:fig}, we
note PV Cep shows a significant variability (in a period of 2-3
years) during which an outburst has occurred, whose declining
phase ($\Delta$J $\approx$ 3 mag) lasted about 120 days. Such a
time interval appears indeed short when compared with the classical
declining times toward the quiescence exhibited by the EXor objects (typically months/years).
Nevertheless, the occurrence of intermittent accretion events from
the circumstellar disk onto the central star, that are widely
accepted as the origin of EXor outbursts, cannot be ruled out just
on the basis of the light-curve shape.

The near-infrared color variations of PVCep are
similar to those of EXor, that tend to be bluer when outbursting and
redder when fading. This behaviour is likely due to the
circumstance that when outbursting the central star becomes more
visible, while when declining the disk dominates. As a result the
near-IR colors related to accretion outbursts and those related to
extinction variations have the same appearance.

At variance with other EXor objects (Lorenzetti et al 2009),
the global behaviour from V to K bands indicates that PV Cep variations
occur behind a not negligible foreground extinction (A$_V$
$\sim$ 5-7 mag). The accretion disk responsible for the near-infrared color
variations does not influence the variations of the optical colors.
On the contrary, the irregular fluctuations of these latter appear mainly
influenced by scattering contributions due to the variable nebula.
Possible contributions from small extinction variations and from the
presence of hot spots could affect just the optical regime by
provoking some {\it noise} in the corresponding colors.

The occurrence of accretion events is also supported by an observational
fact that cannot be accounted for in a scheme of pure extinction
variations. This fact is the time lag in the light-curves discussed in Sect.3.2.
An extinction variation in fact, cannot determine any time delay
between the optical and the near-IR outburst since it acts as a
mere screen able to produce only simultaneous changes at any
wavelength.

About the PV Cep classification, we conclude that it
presents many differences with respect to the known EXor and we
believe it is an object more massive and more complex than other
EXor. The case of PV Cep could lead to argue about the
classification of some recently discovered young sources in
outburst, that are referred, maybe over-simplifying, as EXor.

Noticeably, some of our conclusions, although reached in a
completely different way, substantially agree with those traced by
Kun et al. (2011). These concern the important role of accretion,
the origin of the nebula variations, the presence of a not
homogeneous components in the inner disk, the fact PV Cep differs
from EXor type stars; on the contrary, their hypothesis of a
variable extinction remains controversial.

\section{Acknowledgements}

The authors would like to thank Fabrizio Massi for helpful
discussions on scattering properties and Riccardo Leoni for the
support given during part of the observations at Campo Imperatore.

{}

\begin{deluxetable}{lccccccc}
\tabletypesize{\footnotesize} \tablecaption{Polarization
measurements of PV Cep. \label{polar:tab}} \tablewidth{0pt}
\tablehead{Date       &  $\lambda$ & P   & $\sigma_P$        & PA   & $\sigma_{PA}$    & apert. & ref \\
           &    (\AA)   & \multicolumn{2}{c}{(\%)}&\multicolumn{2}{c}{(deg)}&(arcsec)&    }
 \startdata
1980 Oct 23& 1.25$\mu$m  & 11.9   &   1.1    & 101.4  &  --- & --- & 1  (10.38 mag) \\
1980 Oct 23& 1.65$\mu$m  &  7.7   &   0.3    & 106.6  &  --- & --- & 1  (8.48 mag)  \\
1980 Oct 23& 2.20$\mu$m  &  3.9   &   0.1    & 107.2  &  --- & --- & 1  (6.73 mag)  \\
1981 Jun 21& 1.25$\mu$m  & 14.3   &   1.2    & 109.2  &  --- & --- & 1  (10.76 mag) \\
1981 Jun 21& 1.65$\mu$m  &  9.1   &   0.2    & 111.4  &  --- & --- & 1  (8.72 mag)  \\
1981 Jun 21& 2.20$\mu$m  &  5.1   &   0.1    & 113.0  &  --- & --- & 1  (6.77 mag)  \\
1984 Jun   & 4000-10000  &  9.7   &   0.9    &   71   &   3   &  6  & 2  \\
1986 Aug 5 & 7675        & 15.2   &   0.6    &  75.3  &  1.2  &8.3  & 3  \\
1986 Aug 10& 7675        & 11.8   &   0.5    &  76.8  &  1.2  &8.3  & 3  \\
1986 Aug 12& 7675        & 14.6   &   0.4    &  76.3  &  1.0  &8.3  & 3  \\
1986 Aug 13& 4700        & 11.9   &   1.1    &  75.9  &  2.6  &8.3  & 3  \\
1986 Aug 12& 7675        & 15.2   &   0.5    &  77.6  &  1.0  &8.3  & 3  \\
1989 Jul   & 4000-10000  & 16.2   &   0.3    &  89.9  &  0.5  &  6  & 2  \\
\enddata
\tablenotetext{a}{~~References to the Table: (1) Lacasse 1982; (2)
Scarrott et al. 1991; (3) M\'{e}nard \& Bastien 1992.}

\end{deluxetable}

\begin{deluxetable}{ccccccccccccc}
\tabletypesize{\scriptsize} \tablewidth{0pt}
\tablecaption{Observed Log of our observations of PV
Cep.\label{summary:tab}} \tablehead{ \colhead{Date} & \colhead{B}
& \colhead{V} & \colhead{R} & \colhead{I} & \colhead{J} &
\colhead{H} & \colhead{K} & \colhead{P$_R$} & \colhead{P$_I$} &
\colhead{PA$_R$} & \colhead{PA$_I$} & \colhead{$\theta_{nebula}$} }
\startdata
(JD-2450000)&    \multicolumn{7}{c}{(mag)}   &\multicolumn{2}{c}{(\%)} &\multicolumn{3}{c}{(deg)}\\
\cline{1-13}
4177        &       &       &       &        &       &  9.32 &       &     &    &    &    &      \\
4229        &       &       &       &        & 11.51 &  9.43 &       &     &    &    &    &      \\
4241        &       &       &       &        & 11.52 &  9.33 &  7.54 &     &    &    &    &      \\
4247        &       &       &       &        & 11.16 &  9.12 &  7.41 &     &    &    &    &      \\
4259        &       &       &       &        & 11.20 &  9.07 &  7.29 &     &    &    &    &      \\
4261        &       &       &       &        & 10.90 &  8.90 &  7.24 &     &    &    &    &      \\
4267        & 19.86 & 17.67 & 15.95 &  13.94 & 10.82 &  8.90 &  7.31 &     &    &    &    & -22  \\
4270        & 20.35 & 17.64 & 16.06 &  14.21 &       &       &       &     &    &    &    & -23  \\
4272        &       &       &       &        & 11.00 &  8.94 &  7.23 &     &    &    &    &      \\
4280        &       &       &       &        & 10.73 &  8.75 &  7.15 &     &    &    &    &      \\
4284        &       &       &       &        & 11.25 &  8.99 &  7.21 &     &    &    &    &      \\
4313        & 19.88 & 18.25 & 16.50 &        &       &       &       &     &21.4&    &68.4& -52  \\
4317        &       & 18.31 & 16.73 &        &       &       &       &     &22.2&    &70.0& -54  \\
4319        &       & 18.20 & 16.75 &        &       &       &       &     &21.2&    &65.7& -52  \\
4329        &       & 18.36 & 16.80 &  15.12 &       &       &       &     &    &    &    & -45  \\
4333        &       & 18.10 & 16.66 &  14.98 &       &       &       &     &22.7&    &66.0& -37  \\
4341        &       & 18.11 & 16.76 &        &       &       &       &     &23.0&    &71.6& -31  \\
4342        &       & 18.64 & 16.89 &  15.05 &       &       &       &     &24.7&    &73.6& -32  \\
4343        &       & 18.04 & 16.57 &  14.93 &       &       &       &     &20.6&    &66.2& -18  \\
4344        &       & 18.10 & 16.55 &  14.91 &       &       &       &     &18.9&    &68.6& -11  \\
4345        &       & 18.48 & 16.48 &  14.89 &       &       &       &     &18.9&    &67.4& - 9  \\
4346        &       & 19.08 & 16.69 &  14.98 &       &       &       &     &22.7&    &72.7& -20  \\
4348        &       & 18.27 & 16.50 &  15.06 &       &       &       &10.2$^*$& &58.8$^*$&& -15  \\
4351        &       &       & 16.79 &  15.24 &       &       &       &23.1$^*$& &71.7$^*$&& - 8  \\
4352        &       &       & 16.85 &  15.17 &       &       &       &18.0$^*$& &71.9$^*$&& -12  \\
4355        & 20.63 & 18.10 & 16.54 &  14.93 &       &       &       &15.1$^*$&17.5&72.2$^*$&68.0& - 8 \\
4358        &       &       & 16.51 &  14.86 &       &       &       &        &19.4&    &70.8& - 6  \\
4360        &       & 18.67 & 16.53 &  14.70 &       &       &       &     &20.0&    &66.0&  0   \\
4361        &       & 18.04 & 16.48 &  14.71 &       &       &       &     &17.7&    &67.0& - 3  \\
4363        &       & 17.95 & 16.29 &  14.33 &       &       &       &     &21.8&    &63.6& - 3  \\
4364        & 19.52 & 18.02 & 16.45 &  14.46 &       &       &       &     &21.4&    &63.8& - 3  \\
4369        &       & 18.41 & 16.50 &  14.68 &       &       &       &18.4$^*$& &65.8$^*$&& - 9  \\
4373        &       & 18.08 & 16.54 &  14.66 &       &       &       &18.4$^*$& &66.8$^*$&& -10  \\
4374        &       & 18.24 & 16.66 &  14.55 &       &       &       &15.5$^*$& &60.8$^*$&& - 7  \\
4375        &       &       & 16.45 &        &       &       &       &18.6$^*$& &61.1$^*$&& - 3  \\
4382        &       &       &       &        & 11.59 &  9.25 &  7.30 &     &    &    &    &      \\
4384        &       &       & 16.51 &        &       &       &       &19.4&22.6&74.9&68.6 & - 2  \\
4385        &       &       &       &        & 11.39 &  9.12 &  7.24 &    &    &    &     &      \\
4386        &       &       & 16.53 &        &       &       &       &16.7&23.2&67.2&67.2 & - 3  \\
4388        &       &       &       &        & 11.41 &  9.05 &  7.12 &    &    &    &     &      \\
4389        &       &       & 16.58 &        & 11.53 &  9.13 &  7.27 &19.0&19.3&69.6&68.9 & - 3  \\
4390        &       &       &       &        & 11.47 &  9.05 &  7.21 &    &    &    &     &      \\
4395        &       & 17.92 & 16.37 &  14.42 &       &       &       &22.5$^*$&&65.5$^*$& & - 8  \\
4398        &       &       &       &        & 11.07 &  8.86 &  7.09 &    &    &   &      &      \\
4401        &       &       &       &        & 11.14 &  8.88 &  7.11 &    &    &   &      &      \\
4403        &       &       &       &        & 11.30 &  9.02 &  7.23 &    &    &   &      &      \\
4405        &       &       & 16.49 &        &       &       &       &    &28.8&   &76.3  & - 1  \\
4409        &       &       &       &        & 11.50 &  9.00 &  7.10 &    &    &   &      &      \\
4410        &       & 18.13 & 16.02 &  14.54 &       &       &       &14.9$^*$&&70.7$^*$& & - 5  \\
4422        &       & 17.61 & 16.12 &  14.56 &       &       &       &24.3$^*$&&79.5$^*$& & - 7  \\
4447        &       &       & 15.85 &        &       &       &       &22.2$^*$&&69.0$^*$& & -11  \\
4455        &       &       & 15.82 &  13.92 &       &       &       &21.8$^*$&&65.5$^*$& & -14  \\
4468        &       &       & 15.32 &  13.77 &       &       &       &    &    &   &      & -13  \\
4469        &       &       & 15.06 &  13.53 &       &       &       &    &    &   &      & -15  \\
4480        &       &       & 16.06 &  13.83 &       &       &       &    &    &   &      & -11  \\
4501        &       &       &       &  13.38 &       &       &       &    &    &   &      & -15  \\
4522        &       & 17.52 & 15.94 &  14.23 &       &       &       &    &17.2&   &60.5  & -25  \\
4557        &       &       &       &        & 10.27 &  8.29 &  6.70 &    &    &   &      &      \\
4558        &       & 17.24 & 15.43 &  13.35 & 10.09 &  8.12 &  6.54 &22.4$^*$&&52.7$^*$& & -18  \\
4559        &       &       & 15.45 &  13.36 &       &       &       &19.2$^*$&&48.2$^*$& & -18  \\
4561        &       &       & 15.18 &  13.24 &       &       &       &    &    &   &      & -25  \\
4564        &       &       &       &        &  9.98 &  8.16 &  6.46 &    &    &   &      &      \\
4569        &       &       &       &        & 10.14 &  8.13 &  6.57 &    &    &   &      &      \\
4576        &       &       & 15.76 &        &       &       &       &    &    &   &      & -30  \\
4580        &       & 17.32 & 15.78 &  13.74 &       &       &       &17.3$^*$&&51.1$^*$& & -29  \\
4583        &       &       & 15.95 &  13.93 &       &       &       &24.7$^*$&&35.8$^*$& & -24  \\
4590        &       &       & 15.94 &  14.12 &       &       &       &22.8$^*$&&43.7$^*$& & -30  \\
4592        &       & 17.39 & 15.85 &        &       &       &       &    &    &   &      & -37  \\
4595        &       &       &       &        & 10.85 &  8.54 &  6.76 &    &    &   &      &      \\
4612        &       & 17.59 & 16.26 &  14.56 &       &       &       &    &    &   &      & -32  \\
4613        &       &       &       &        & 11.36 &  8.93 &  7.08 &    &    &   &      &      \\
4614        &       &       &       &        & 11.45 &  9.01 &  7.08 &    &    &   &      &      \\
4623        &       & 17.94 & 16.67 &  14.97 &       &       &       &    &    &   &      & -46  \\
4627        &       &       &       &        & 11.96 &  9.36 &  7.38 &    &    &   &      &      \\
4630        &       &       &       &        & 12.27 &       &       &    &    &   &      &      \\
4631        &       &       &       &        & 12.36 &       &       &    &    &    &      &      \\
4633        &       &       &       &        & 12.34 &  9.71 &  7.62 &    &    &    &      &      \\
4634        &       &       &       &        & 12.35 &  9.73 &  7.68 &    &    &    &      &      \\
4636        &       &       &       &        & 12.34 &  9.74 &  7.73 &    &    &    &      &      \\
4637        &       &       &       &        & 12.40 &  9.79 &  7.72 &    &    &    &      &      \\
4639        &       & 18.23 & 16.64 &  15.00 & 12.19 &  9.63 &  7.58 &    &16.6&    &54.1  & -49  \\
4641        &       &       &       &        & 12.23 &  9.62 &  7.54 &    &    &    &      &      \\
4642        &       &       &       &        & 12.28 &  9.62 &  7.53 &    &    &    &      &      \\
4643        &       &       &       &        & 12.17 &  9.55 &  7.51 &    &    &    &      &      \\
4644        &       &       &       &        & 12.04 &  9.48 &  7.43 &    &    &    &      &      \\
4645        &       &       &       &        & 12.08 &  9.54 &  7.55 &    &    &    &      &      \\
4646        &       &       &       &        & 12.00 &  9.48 &  7.47 &    &    &    &      &      \\
4648        &       &       &       &        & 12.03 &  9.54 &  7.55 &    &    &    &      &      \\
4649        & 18.81 & 18.36 & 16.83 &  15.11 & 12.02 &  9.54 &  7.56 &13.6&    &64.8&      & -52  \\
4650        &       &       &       &        & 12.06 &  9.54 &  7.50 &    &    &    &      &      \\
4651        &       &       &       &        & 12.12 &  9.59 &  7.54 &    &    &    &      &      \\
4652        &       &       &       &        & 12.21 &  9.64 &  7.57 &    &    &    &      &      \\
4653        &       &       &       &        & 12.36 &  9.79 &  7.61 &    &    &    &      &      \\
4654        &       & 18.35 & 17.06 &  15.22 & 12.37 &  9.81 &  7.67 &14.3&    &64.4&      & -53  \\
4655        &       &       &       &        & 12.26 &  9.69 &  7.47 &    &    &    &      &      \\
4659        &       &       &       &        & 12.06 &  9.62 &  7.61 &    &    &    &      &      \\
4663        &       & 18.74 & 17.01 &  15.34 &       &       &       &    &    &    &      & -51  \\
4664        &       &       &       &        & 12.52 &  9.96 &  7.85 &    &    &    &      &      \\
4666        & 20.25 &       &       &        &       &       &       &    &    &    &      &      \\
4667        &       & 18.41 & 16.94 &  15.35 &       &       &       &18.9&    &61.7&      & -51  \\
4672        &       &       &       &        & 12.66 & 10.04 &  7.80 &    &    &    &      &      \\
4673        &       &       &       &        & 12.61 & 10.00 &  7.79 &    &    &    &      &      \\
4676        &       &       &       &        & 12.81 & 10.20 &  7.94 &    &    &    &      &      \\
4677        &       & 18.16 & 16.89 &  15.22 &       &       &       &12.0&    &60.0&      & -48  \\
4680        &       & 18.46 & 16.86 &  15.20 &       &       &       &14.5&    &62.9&      & -50  \\
4682        &       & 18.39 & 16.94 &  15.30 &       &       &       &13.2&    &62.9&      & -49  \\
4685        &       &       &       &        & 12.68 & 10.03 &  7.70 &    &    &    &      &      \\
4690        &       &       &       &        & 12.68 & 10.03 &  7.73 &    &    &    &      &      \\
4691        &       & 18.57 & 16.93 &  15.28 &       &       &       &    &16.6&    &54.0  & -46  \\
4692        &       &       &       &        & 12.71 & 10.05 &  7.77 &    &    &    &      &      \\
4697        &       &       &       &        & 12.75 & 10.11 &  7.88 &    &    &    &      &      \\
4698        &       &       &       &        & 12.80 & 10.18 &  7.98 &    &    &    &      &      \\
4701        &       &       &       &        & 12.75 & 10.19 &  7.87 &    &    &    &      &      \\
4705        &       &       &       &        & 12.77 & 10.21 &  7.80 &    &    &    &      &      \\
4708        &       & 18.19 & 16.65 &        & 12.74 & 10.21 &  7.88 &     &16.2&    &66.7  & -48  \\
4711        &       &       &       &        & 12.45 &  9.90 &  7.59 &     &    &    &      &      \\
4713        &       &       &       &        & 12.47 &  9.85 &  7.52 &     &    &    &      &      \\
4714        &       &       &       &        & 12.50 &  9.91 &  7.68 &     &    &    &      &      \\
4716        &       &       &       &        & 12.40 &  9.86 &  7.61 &     &    &    &      &      \\
4719        &       &       &       &        & 12.51 &  9.98 &  7.68 &     &    &    &      &      \\
4720        &       & 17.79 & 16.46 &  14.85 & 12.47 &  9.94 &  7.68 &     &    &    &      & -39  \\
4723        &       & 17.94 & 16.49 &  15.04 &       &       &       &     &    &    &      & -47  \\
4724        &       &       & 16.60 &  14.87 & 12.54 & 10.01 &  7.79 &     &16.1&    &63.1  & -50  \\
4727        &       &       & 16.60 &  14.87 & 12.47 &  9.91 &  7.65 &     &    &    &      &      \\
4728        &       &       & 16.60 &  14.87 & 12.47 &  9.92 &  7.70 &     &    &    &      &      \\
4733        &       &       & 16.54 &  15.00 &       &       &       &     &    &    &      & -46  \\
4734        &       & 17.76 & 16.38 &  14.82 &       &       &       &     &15.7&    &65.1  & -48  \\
4737        &       & 17.86 & 16.47 &        &       &       &       &     &13.2&    &60.7  & -51  \\
4738        &       & 17.83 & 16.46 &  14.84 & 12.28 &  9.84 &  7.67 &     &14.5&    &61.3  & -48  \\
4745        &       & 17.60 & 16.42 &  14.76 &       &       &       &     &    &    &      & -49  \\
4746        &       & 17.92 & 16.36 &  14.90 &       &       &       &     &    &    &      & -48  \\
4748        &       & 17.75 & 16.39 &  14.79 &       &       &       &     &    &    &      & -51  \\
4749        &       & 17.81 & 16.33 &  14.78 &       &       &       &     &18.5&    &63.4  & -49  \\
4750        &       & 17.89 & 16.41 &        &       &       &       &     &17.4&    &63.2  & -48  \\
4752        &       & 17.55 & 16.22 &  14.68 &       &       &       &     &16.9&    &62.7  & -45  \\
4754        &       &       & 16.45 &  14.91 & 12.23 &  9.78 &  7.63 &     &    &    &      & -47  \\
4757        &       &       &       &        & 12.40 &  9.99 &  7.86 &     &    &    &      &      \\
4759        &       &       &       &        & 12.22 &  9.78 &  7.64 &     &    &    &      &      \\
4760        &       &       &       &        & 12.21 &  9.77 &  7.68 &     &    &    &      &      \\
4762        &       & 17.66 & 16.32 &  14.76 &       &       &       &15.3 &    &54.8&      & -48  \\
4765        &       &       &       &        & 12.09 &  9.67 &  7.43 &     &    &    &      &      \\
4766        &       & 17.92 & 16.32 &  14.76 & 12.13 &  9.71 &  7.68 &     &17.1&    &64.1  & -49  \\
4769        &       & 17.87 & 16.44 &  14.83 &       &       &       &     &15.8&    &63.6  & -45  \\
4772        &       &       &       &        & 12.25 &  9.87 &  7.78 &     &    &    &      &      \\
4773        &       &       &       &        & 12.12 &  9.71 &  7.66 &     &    &    &      &      \\
4774        &       & 17.90 & 16.35 &  14.75 &       &       &       &     &19.3&    &63.4  & -43  \\
4779        &       &       &       &        & 12.04 &  9.66 &  7.61 &     &    &    &      &      \\
4780        &       &       &       &        & 12.20 &  9.81 &  7.80 &     &    &    &      &      \\
4781        &       &       &       &        & 12.16 &  9.83 &  7.81 &     &    &    &      &      \\
4782        &       &       &       &        & 12.13 &  9.83 &  7.85 &     &    &    &      &      \\
4788        &       &       & 16.53 &  14.86 &       &       &       &     &    &    &      & -41  \\
4790        &       &       &       &        & 11.86 &  9.73 &  7.80 &     &    &    &      &      \\
4804        &       & 17.81 & 16.40 &  14.77 &       &       &       &19.6 &    &65.5&      & -34  \\
4833        &       &       & 16.61 &  15.05 &       &       &       &     &    &    &      & -30  \\
4836        &       &       & 16.82 &  15.12 &       &       &       &19.8$^*$&&64.4$^*$&   & - 8  \\
4839        &       &       & 16.78 &  15.17 &       &       &       &23.6$^*$&&68.9$^*$&   & -11  \\
4863        &       &       &       &  15.48 &       &       &       &     &    &    &      &      \\
4866        &       &       & 17.78 &  15.59 &       &       &       &     &    &    &      &      \\
4884        &       &       &       &        & 12.73 & 10.24 &  7.94 &     &    &    &      &      \\
4897        &       &       & 17.05 &  15.37 &       &       &       &     &    &    &      &      \\
4903        &       &       &       &        & 12.50 &  9.98 &  7.75 &     &    &    &      &      \\
4904        &       &       &       &        & 12.58 & 10.10 &  7.87 &     &    &    &      &      \\
4906        &       &       & 17.47 &        &       &       &       &     &    &    &      &      \\
4908        &       &       &       &        & 12.23 &  9.75 &  7.58 &     &    &    &      &      \\
4910        &       &       &       &  15.09 &       &       &       &19.0$^*$&&69.7$^*$&   &      \\
4913        &       &       &       &        & 12.51 & 10.05 &  7.80 &     &    &    &      &      \\
4923        &       &       & 16.97 &  15.29 &       &       &       &     &    &    &      &      \\
4930        &       &       & 16.98 &  15.17 &       &       &       &     &    &    &      &      \\
4935        &       &       &       &  15.43 &       &       &       &     &    &    &      &      \\
4942        &       &       & 16.92 &  15.25 &       &       &       &     &    &    &      &      \\
4944        &       &       & 17.15 &  15.26 &       &       &       &     &14.0&    &61.9  & - 6  \\
4945        &       &       & 16.82 &  15.10 &       &       &       &     &    &    &      & - 2  \\
4946        &       & 18.69 & 16.94 &  15.22 &       &       &       &     &15.0&    &61.4  & -10  \\
4948        &       & 18.83 & 16.95 &  15.33 &       &       &       &     &16.0&    &67.4  &   2  \\
4953        &       &       & 17.14 &  15.24 &       &       &       &     &    &    &      &      \\
4959        &       &       & 17.39 &  15.34 &       &       &       &     &    &    &      &      \\
4960        &       &       & 17.16 &  15.36 &       &       &       &11.8$^*$&&69.0$^*$&   &      \\
4973        &       & 18.56 & 17.02 &  15.27 &       &       &       &     &13.9&    &69.0  & - 4  \\
4993        &       & 18.70 & 17.21 &  15.40 &       &       &       &     &20.5&    &65.1  & - 2  \\
4995        &       & 19.18 & 17.21 &  15.55 &       &       &       &     &    &    &      &      \\
5010        &       & 18.89 & 17.16 &  15.49 &       &       &       &     &    &    &      & - 4  \\
5011        &       & 18.90 & 17.08 &  15.42 &       &       &       &     &19.5&    &63.4  &   5  \\
5012        &       & 18.83 & 17.20 &  15.44 &       &       &       &     &18.4&    &63.3  & - 9  \\
5020        &       & 18.73 & 17.20 &  15.40 &       &       &       &     &23.7&    &66.1  &   0  \\
5026        &       & 18.71 & 17.13 &  15.42 &       &       &       &     &17.9&    &65.0  &   3  \\
5035        & 19.99 & 18.77 &       &  15.39 &       &       &       &14.9 &    &63.5&      & - 5  \\
5038        &       &       & 17.13 &  15.45 &       &       &       &     &19.0&    &61.1  & - 4  \\
5048        &       &       & 17.22 &  15.67 &       &       &       &     &22.3&    &59.8  & - 3  \\
5053        &       &       & 17.31 &  15.54 &       &       &       &     &20.2&    &63.7  & - 4  \\
5062        &       &       & 17.45 &  15.77 &       &       &       &     &20.6&    &62.7  & - 1  \\
5065        &       &       & 17.76 &  15.95 &       &       &       &     &24.2&    &62.2  & - 5  \\
5068        &       &       & 17.73 &  15.94 &       &       &       &     &21.2&    &57.3  & - 1  \\
5080        &       &       & 17.16 &  15.19 &       &       &       &     &21.6&    &58.6  & - 5  \\
5082        &       &       & 17.34 &  15.28 &       &       &       &     &22.0&    &53.8  & - 9  \\
5091        &       &       & 17.04 &  15.01 &       &       &       &     &19.6&    &55.2  & -12  \\
5095        &       &       & 17.01 &  15.12 &       &       &       &     &22.1&    &56.8  & - 8  \\
5107        &       & 19.27 & 17.25 &  15.39 &       &       &       &     &17.9&    &68.2  & - 3  \\
5113        &       &       & 17.48 &  15.41 &       &       &       &     &    &    &      &      \\
5123        &       &       & 17.48 &  15.41 &       &       &       &     &19.3&    &57.6  &      \\
5135        &       &       & 17.48 &  15.41 &       &       &       &     &24.5&    &52.9  &      \\
5150        &       &       & 17.48 &  15.41 &       &       &       &     &23.0&    &58.7  & -14  \\
5153        &       &       & 17.48 &  15.41 &       &       &       &     &18.5&    &58.2  & - 8  \\
\cline{1-13}
\cline{1-13}
\enddata
\tablenotetext{a}{Typical errors of the optical (near/IR) magnitude never
exceed 0.02 mag, while errors on P and PA are of the order of 1-3
\% and 2-4$^{\circ}$, respectively. The uncertainty of
$\theta_{nebula}$ is about 1-2$^{\circ}$.}
\tablenotetext{b}{Polarization and PA values obtained in unfiltered mode.}
\end{deluxetable}

\end{document}